\documentclass{aa}  

\usepackage{placeins}

\usepackage{natbib}
\bibpunct{(}{)}{;}{a}{}{,} 
\usepackage{graphicx}
\usepackage{newtxtext,newtxmath}
\usepackage{lscape}
\usepackage{subcaption}
\usepackage{gensymb}
\usepackage{siunitx}
\DeclareSIUnit[]{\angstrom}{\textup{\AA}}
\DeclareSIUnit[]{\Rsun}{\text{\ensuremath{R_{\sun}}}}
\DeclareSIUnit[]{\Msun}{\text{\ensuremath{M_{\sun}}}}

\usepackage{pgfplots}
\usepackage{color}

\newcommand{\unsim}{\mathord{\sim}}

\usepackage[]{hyperref}
\hypersetup{
  colorlinks   = true, 
  urlcolor     = blue, 
  linkcolor    = blue, 
  citecolor   = blue 
}
\begin{document}

   \title{Cyclotron emitting magnetic white dwarfs in post common envelope binaries discovered with the Zwicky Transient Facility}


   \author{J. van Roestel \inst{1} \and
          A.C.~Rodriguez \inst{2} \and
          P.~Szkody \inst{3} \and \\
          A.J.~Brown \inst{4} \and 
          I.~Caiazzo  \inst{2} \and
          A.~Drake  \inst{2} \and
          K.~El-Badry  \inst{2} \and
          T.~Prince  \inst{2} \and 
          R.M.R.~Rich  \inst{5} \and
          J.D.~Neill  \inst{6} \and
          Z.~Vanderbosch  \inst{2} \and \\
          E.C.~Bellm \inst{6} \and
          R.~Dekany  \inst{7} \and
          F.~Feinstein \inst{8} \and
          M.~Graham  \inst{2} \and  
          S.L.~Groom  \inst{9} \and
          G.~Helou \inst{9} \and
          S.R.~Kulkarni \inst{2} \and
          T.~du Laz \inst{2} \and
          A.~Mahabal  \inst{2} \and 
          Y.~Sharma \inst{2} \and
          J.~Sollerman  \inst{10} \and
          A.~Wold  \inst{9} 
          }
   \institute{Anton Pannekoek Institute for Astronomy, University of Amsterdam, 1090 GE Amsterdam, The Netherlands
         \and Division of Physics, Mathematics, and Astronomy, California Institute of Technology, Pasadena, CA 91125, USA
        \and Department of Astronomy, University of Washington, Seattle, WA 98195, USA
        \and Departament de Física, Universitat Politècnica de Catalunya, c/Esteve Terrades 5, 08860 Castelldefels, Spain
        \and Department of Physics \& Astronomy, Univ. of California Los Angeles, PAB 430 Portola Plaza, Los Angeles, CA 90095-1547, USA
        \and Caltech Optical Observatories, California Institute of Technology, Pasadena, CA  91125
        \and DIRAC Institute, Department of Astronomy, University of Washington, 3910 15th Avenue NE, Seattle, WA 98195, USA 
        \and Aix Marseille University, CNRS/IN2P3, CPPM, Marseille, France
        \and IPAC, California Institute of Technology, 1200 E. California Blvd, Pasadena, CA 91125, USA
        \and The Oskar Klein Centre, Department of Astronomy, Stockholm University, AlbaNova, SE-10691 Stockholm, Sweden
        }

   \date{Received November, 2024; accepted 12 12, 2024}

\titlerunning{Low accretion rate magnetic white dwarfs in PCEBs found with ZTF}

 
  \abstract
    {We present the discovery of 14 new (and recovery of 4 known) low accretion rate magnetic white dwarfs in post-common envelope binaries that emit strong cyclotron emission using the Zwicky Transient Facility (ZTF) light curves, doubling the known sample size. In addition, we discovered a candidate magnetic period bouncer and recovered three known ones.
    We confirmed the presence of cyclotron emission using low-resolution spectra in 19 objects.  Using the ZTF light curves, follow-up spectra, and the spectral energy distribution, we measured the orbital period, magnetic field strength, and white dwarf temperature of each system. Although the phase-folded light curves have diverse shapes and show a much larger variability amplitude, we show that their intrinsic properties (e.g. period distribution, magnetic field strength) are similar to those of previously known systems. The diversity in light curve shapes can be explained by differences in the optical depth of the accretion spot and geometric differences, the inclination angle and the magnetic spot latitude.
    The evolutionary states of the longer period binaries are somewhat uncertain but are vary; we found systems consistent with being pre-polars, detached polars, or low-state polars. In addition, we discovered two new low-state polars that likely have brown dwarf companions and could be magnetic period bouncers.} 
   \maketitle

%

\section{Introduction}\label{sec:intro}
Most white dwarfs with close low-mass binary companions (K, M, or brown dwarf stars) are binary systems that evolved through a common envelope phase \citep{paczynski1976} and are called post-common envelope binaries (PCEBs). During the very short-lived common envelope phase, the shared envelope is ejected while the stars rapidly spiral inward. 
After these binaries emerge from the common envelope, angular momentum losses slowly decrease the orbital period further \citep{paczynski1967,verbunt1981}. When the companion star fills its Roche lobe, stable Roche lobe overflow accretion begins, and the system becomes a cataclysmic variable \citep[CV; see][]{warner1995}. As CVs, these binaries keep losing angular momentum, and their orbital periods slowly evolve over time. They initially have high accretion rate and their orbital period is decreasing until they reach the period minimum at $\approx$80 minutes \citep{gansicke2009} and their period increases again but with even lower accretion rates. For an in-depth review of CV evolution, we refer the reader to \citet{knigge2011}. 

White dwarfs can have strong magnetic fields, from a few kG to hundreds of \unit{MG} \citep{wickramasinghe2000,landstreet2019,bagnulo2020}. 
One of the outstanding issues is how magnetic fields can change the evolution and appearance of PCEBs, but also whether magnetic fields are generated as a result of binary interactions like mass transfer. Observations show that if a white dwarf has strong magnetic fields ($\gtrsim \qty{1}{MG}$), the accretion flow in a CV is partially or completely disrupted, and the CV forms only a ring-like structure or no disk at all \citep{ferrario2015}. Instead, the accretion flow is redirected along the magnetic-field lines and impacts the white dwarf on the magnetic poles. These systems are called `polars' (or AM~Her-type) or `intermediate polars' (or DQ~Her-type), see \citet{cropper1990}. They constitute a significant part of the CV population; 20-25\% \citep{ferrario2015}, or possibly as many as 36\% \citep{pala2020}. As was already noted in \citet{liebert2005}, a much lower occurrence rate of magnetic systems is observed in detached PCEBs/pre-CVs suggesting that magnetic fields are generated at some point during PCEB or CV evolution. \citet{tout2008} suggested that the magnetic field was generated during the common envelope phase, but this does not agree with the observed population \citep{belloni2020}.

Although rare, there is a small and growing number of magnetic white dwarfs in PCEBs that do not show any Roche lobe overflow accretion (the typical pre period-minimum Roche lobe accretion rate is $\dot{\mathrm{M}} \gtrsim 10^{-10.5}$ \unit{M_\odot yr^{-1})} , but instead feature wind accretion ($\dot{\mathrm{M}} \lesssim10^{-13}$ \unit{M_\odot yr^{-1}}, \citealt{schwope2002,schmidt2005,webbink2005,schwope2009,parsons2021}. These accretion rates are consistent with the estimated wind mass-loss rate for M-dwarfs $10^{-12}$--$10^{-15}$ \unit{M_\odot yr^{-1}} 
\citep[e.g.][]{johnstone2015,wood2021}. 
Observationally, these systems show (strong) cyclotron emission, often at optical wavelengths. Cyclotron emission is a result of charged non-relativistic particles moving through a strong magnetic field and is a dominant effect at very low accretion rates ($\dot{\mathrm{M}} \approx10^{-13}$ \unit{M_\odot yr^{-1}},  \citealt{wickramasinghe2000,szkody2003}). The first two of such systems (WX~LMi and HS~0922+1333) were found by spectroscopic follow-up of blue sources identified in the Hamburg Quasar Survey \citep{reimers1999,reimers2000}. Additional systems were found using SDSS spectra \citep{szkody2003,schmidt2005}. A recent overview of most of the 17 currently known candidate systems is given by \citet[][Table~2]{parsons2021}. 

These systems were initially named Low Accretion Rate Polars \citep[LARPs;][]{schwope2002} because polars can exhibit prolonged low states with significantly lower accretion rates than expected for Roche lobe overflow. For example, EF~Eri is a polar that switched to a low state in 1997 and remained in that state until the end of 2023. However, \citet[][]{schwope2009} showed that most LARPs were not low-state polars, but pre-polars (PrePs) instead; the donors were far from filling their Roche lobes. 

The most recent addition to the family of detached magnetic white dwarfs in PCEBs are white dwarf pulsars: AR~Sco \citet{marsh2016} and J191213.72-441045.1 \citep{pelisoli2023,schwope2023}. In these systems, the white dwarf spin period is not synchronised with the orbit, resulting in pulsar-like behaviour because of synchrotron emission, which is thought to be from the interaction of the white dwarf magnetic field with that of the donor star magnetic field \citep[e.g.][]{marcote2017,duplessis2022} and no cyclotron emission as in the other systems. 

Recently, \citet{schreiber2022} suggested a theory on how magnetic white dwarfs in PCEBs are generated in order to explain the different types of systems observed. The proposed theory includes a long detached phase in between stages of Roche lobe mass transfer, which would be another possible evolutionary state.

There is also another (almost equally rare) category of low accretion rate magnetic white dwarfs in binaries. \citet{breedt2012} and \citet{kawka2021} identified four magnetic white dwarfs with brown dwarf companions in period of $\approx$ 90 minutes (see also Table B.1 in \citealt{schreiber2023} for an overview of candidate magnetic period bouncers).   They are possibly magnetic period bouncers; systems where the donor has been stripped of most mass and the Roche lobe mass loss rate is $\dot{\mathrm{M}} \lesssim 10^{-11}$ \unit{M_\odot yr^{-1})}, similar to wind-accretion mass-transfer rates Even though these systems are (likely) still filling their Roche lobe and should therefore considered to be a sub-class polars, the accretion rate on the same order of magnitude as wind accreting systems and therefore also show strong cyclotron emission. 
Recently, \citet{schreiber2023} suggested that magnetic period bouncers can detach (i.e. the donor is not filling its Roche lobe any more) due to the magnetic field.

To avoid the uncertainty in the evolutionary state of these systems, we use the term ``low-accretion rate magnetic white dwarfs in a PCEB'' as a broad observational category for any system that shows observational properties consistent with a low accretion rate (see Sect. \ref{sec:ZTF}) regardless of their evolutionary state.

To understand the evolutionary history, the magnetic field properties, and how the white dwarf magnetic fields are generated, detailed observations of low accretion rate magnetic white dwarfs in PCEBs are used. \citet{parsons2013a} presented the detailed analysis of eclipsing system SDSS~J030308.35+005444.1 showing that the magnetic field is non-dipolar and that the two magnetic spots are located on the side of the white dwarf facing the companion star. \citet{hakala2022} presented a polarimetry study of seven systems to measure the location, size, shape, and magnetic polarity of the cyclotron-emitting region. They showed that the magnetic field shape of the white dwarf is more complex than a simple dipole. \citet{parsons2021} measured the binary parameters of 6 binary systems. They determined that white dwarfs are more massive compared to detached white~dwarf -- M~dwarf binaries, but more importantly, they confirmed that most of these binaries could have filled their Roche lobes in the past, consistent with the formation channel proposed by \citet{schreiber2022}. \citet{parsons2021} note that many systems do not show obvious signs of magnetic fields in archival spectra but do show anomalous light curves, which could be used to find more of these systems.

In this paper, we present a search for cyclotron emitting PCEBs using the Zwicky Transient Facility (ZTF) light curves. We discovered 14 new pre-polars or low accretion rate polar, doubling the currently known sample of these objects. In addition, we recovered three low-state polars that might be magnetic period bouncers, and identified one new candidate object. In Sect.~\ref{sec:ZTF}, we discuss how we identified these systems in ZTF data and present the ZTF light curves. Sect.~\ref{sec:followup} presents the follow-up spectra that we obtained with various telescopes, and in Sect.~\ref{sec:methods} we briefly discuss the methods used to infer the properties of the binary systems. The results are presented in Sect.~\ref{sec:results}. In Sect.~\ref{sec:discussion}, we discuss the light curve shapes, long time-scale changes in one particular system,  similarities and differences with the previously known sample, and finally we discuss the evolutionary states of the new systems. We look into future prospects in Sect.~\ref{sec:futurework} and summarise the paper in Sect.~\ref{sec:conclusion}.


\section{The Zwicky Transient Facility and target selection}
\label{sec:ZTF}
As part of the ZTF, the Palomar 48-inch (P48) telescope images the sky every night \citep{graham2019,bellm2019,dekany2020}. The exact cadence varies over the course of the survey and area on the sky, but typically the cadence is 1--3 days in each filter. 
Most of the time ZTF uses the $g$ and $r$ bands, but a small fraction of observations are also made in the $i$ band. Exposure times are predominantly 30 seconds for $g$ and $r$ and range from 30 to 90 seconds for $i$-band images. The median limiting magnitude, averaged over the lunar cycle, is \mbox{$\approx20.5$} AB-mag in all three bands. In this work, we used the PSF-fit photometry light curves\footnote{based on the science images, not the difference images} which are automatically generated for all persistent sources detected in the ZTF reference images \citep[for a detailed description, see][]{masci2019}.

 The objects presented in this work were initially identified as anomalous by humans that visually inspected a large sample of ZTF light curves of periodic variable stars for two different projects. Some of the objects were found by humans looking at light curves with the goal of building a training set for a variable star machine learning classifier (\citealt{vanroestel2021,coughlin2021,healy2024}). The second effort, which resulted in most discoveries, focused on periodic variable stars below the main sequence in order to find eclipsing white dwarf -- red dwarf binaries (\citealt{vanroestel2021a}, van Roestel et al. in prep). We briefly discuss the completeness of the search in Sect. \ref{sec:discussion}.

 Using SIMBAD \citep{ochsenbein2000}, we determined that three of the objects were already known: MQ~Dra \citep{szkody2003}, SDSS~J222918.97+185340.2 \citep{drake2014}, and ZTF~J014635.74+491443.1 \citep{guidry2021}. This prompted us to inspect the ZTF light curves of all known low accretion rate magnetic white dwarfs (both PrePs and magnetic period bouncers). These include all systems listed in Table 2 in \citet{parsons2021}, 2MASS J0129+6715 from \citet{krushinsky2020}, and the magnetic period bouncers listed in \citet{kawka2021}\footnote{We do not include EF Eri because it is known to show state changes and therefore we consider it a polar and not a magnetic periodbouncer}.

 With a better understanding of the lightcurve shapes, we again systematically inspected all anomalous periodic ZTF objects and collected objects that show 1) a strictly periodic ZTF light curves that 2) features no irregular variability or state changes (as seen in polars, see recent work with ZTF by \citealt{duffy2022,ok2022}) and 3) show a period-folded light curve that behaved very differently in-between the $g$, $r$, and $i$ bands and/or the light curves exhibited complex and unusual shapes that could be due to cyclotron emission. This resulted in a total of 28 of candidates.
 
 After some further vetting (including the use of spectra, see Sect.~\ref{sec:followup}) we removed 6 ambiguous objects that were most likely just spotted stars. We also inspected CRTS \citep{drake2009} and ATLAS \citep{tonry2018} light curves (that have a longer time baseline) but none of the objects showed state changes typical of those seen in polars. 

Table~\ref{tab:summary} shows 22 objects that we consider strong candidates (discoveries or recoveries) to be low accretion rate magnetic white dwarfs in PCEBs we identified by their ZTF light curves.
The first part of the table lists the 18 candidate LARPs/PrePs/low-state polar candidate (4~previously known) and these are shown in Figure~\ref{fig:lc_spec1}. The second part of the table shows short period systems that are candidate low-state polars or magnetic period bouncers (2 previously known) which are shown in Figure~\ref{fig:lc_spec_polars}. For reference, we also show the ZTF light curves of known objects that we did not recover in Figures \ref{fig:lc_known} and \ref{fig:lc_known_polars} folded on their reported orbital period. In addition, two objects that were previously identified as PrePs/LARPs show state changes and should be reclassified as regular polars are shown in Figure~\ref{fig:lc_known2}. No ZTF data are available for one of the already known systems, HS~0922+1333, because it falls in a chip-gap in the ZTF primary field grid.

\section{Identification spectra}\label{sec:followup}
To confirm that the unusual light curves are indeed the result of cyclotron emission, we obtained an identification spectrum of all objects. Because cyclotron features are very broad, a low-resolution spectrum is, in principle, enough to identify them. Therefore, we first obtained a spectrum for most objects using SEDM (see below). If this spectrum was inconclusive, we used other telescopes to obtain a higher resolution and signal-to-noise spectrum. Table~\ref{tab:summary} lists the spectrographs that were used for each object and the spectra are shown in the right panels of Fig.~\ref{fig:lc_spec1} and Fig.~\ref{fig:lc_spec_polars}. In the rest of this section, we briefly discuss the instrumental setups and the data reduction process.

\subsection{SEDM}
We obtained identification spectra with SEDM for 10 objects. The Spectral Energy Distribution Machine \citep[SEDM;][]{blagorodnova2018} is an integral field spectrograph mounted on the Palomar 60-in telescope. It has a very low resolution (\qty{50}{\angstrom}
 per pixel) and a wavelength range of \qtyrange{3650}{10000}{\angstrom}. The data were automatically reduced using the standard pipeline \citep{rigault2019}.

\subsection{DBSP}
Nine spectra were obtained with the Palomar 200-inch telescope and the Double-Beam Spectrograph \citep[DBSP,][]{oke1982} with a resolution of $R\approx1500$. For each night an average bias and a normalised flat-field frame were made out of 10 individual bias and 10 individual lamp flat fields. Wavelength calibration was done using calibration lamp spectra obtained at the start of the night. For the blue arm, FeAr and for the red arm, HeNeAr arc exposures were taken. Data were reduced using \textsc{DBSP\_DRP}\footnote{\url{https://github.com/finagle29/DBSP_DRP}}, a data reduction pipeline based on \textsc{PypeIt} \citep{prochaska2020}.

\subsection{DIS}
For three targets, we obtained spectra with the \qty{3.5}{m}
telescope at Apache Point Observatory (APO). We used the Double Imaging Spectrograph (DIS) in either a low-resolution (\qty{2}{\angstrom} per pixel) or medium-resolution (\qty{0.6}{\angstrom} per pixel) gratings. The data were reduced using IRAF routines and calibrated using He, Ne, and Ar lamps and flux standards obtained during each night.

\subsection{LRIS}
We obtained three spectra with the \qty{10}{m} Keck\,I Telescope (HI, USA) and the Low Resolution Imaging Spectrometer (LRIS; \citealt{Oke1995,McCarthy1998}).
We used the R600/4000 grism for the blue arm ($R\approx1100$) and the R400 grating for the red arm ($R\approx1000$). The wavelength range is approximately \qtyrange{3200}{10000}{\angstrom}. A standard long-slit data reduction procedure was performed with the \textsc{Lpipe} pipeline\footnote{\url{http://www.astro.caltech.edu/~dperley/programs/lpipe.html}} \citep{Perley2019}.

\subsection{Kast}
We obtained one spectrum with the two-arm Kast spectrograph mounted at the 3-m Shane telescope \citep{miller1994}. In the blue and red arms, we used 600/4310 and 600/5000 gratings. Combined with the \qty{2}{\arcsec} wide slit, the resolution is $R\approx2200$ and $R\approx2500$. We used exposure times between \qtyrange{1500}{3600}{s}, depending on the target brightness. We split the exposures in the red arm to mitigate the effects of cosmic rays. Data were reduced using a \textsc{PypeIt} based pipeline \citep{prochaska2020}.

\subsection{IDS}
We obtained one spectrum with the Intermediate Dispersion Spectrograph (IDS) mounted on the Isaac Newton Telescope (INT). The spectrograph was equipped with the EEV10a CCD and the R400V grating ($R=1452$). Data were reduced using the ASPIRED package\footnote{\url{https://github.com/cylammarco/ASPIRED}} \citep{lam2023}.

\subsection{Sloan Digital Sky Survey}
We used two archival spectra from the Sloan Digital Sky Survey  (SDSS, \citealt{york2000}).
The SDSS is a imaging and spectroscopic survey using the 2.5-m wide-angle optical telescope at Apache Point Observatory in New Mexico, United States. The spectra typically have a total integration time of 45--60 minutes depending on observing conditions. The wavelength coverage of the spectra is continuous from about \qty{3800}{\angstrom} to \qty{9200}{\angstrom} with a resolution of R$\unsim$1800 \citep{Uomoto1999}.

\section{Methods}\label{sec:methods}

\subsection{Light curve analysis}

  For each object, we refined the period using a multiband, multiharmonic Lomb-Scargle method as implemented in \textsc{gatspy} \citep{vanderplas2016}. We used 5 harmonics and used all available ZTF data. Most light curves show a stable periodic signal, but there are a few exceptions that show long timescale (years) changes in the shape of the folded light curve shape such as MQ Dra, ZTF~J1531+0204, and ZTF~J0116+4417 which is discussed further in Sect.~\ref{sec:modulation}.

\subsection{Spectral analysis}
The observational feature that characterises low accretion rate magnetic white dwarfs is the presence of broad cyclotron emission (`humps'), which can be seen in optical, near-infrared, and UV wavelengths, depending on the exact parameters of the system. Investigation of the hydrodynamics of a low accretion flow with a strong magnetic field show that the spot in the atmosphere is heated by a stream of fast ions and cools by emission of cyclotron radiation, the bombardment solution \citep{woelk1996,fischer2001,beuermann2004}. 
The wavelengths of the cyclotron emission peaks depend strongly on the (local) magnetic field strength:
\begin{equation}
    C_n = \frac{\qty{10710}{\angstrom}}{n}\left(\frac{\qty{100}{MG}}{B}\right)
    \label{eq:cyclotron}
\end{equation}
where $n$ indicates the harmonic, $B$ is the magnetic field strength, and $C_n$ is wavelength of the cyclotron harmonic in \r{A}ngstrom. 
The exact appearance of these cyclotron features in the spectrum (shape, strength) depends strongly on the viewing angle and properties of the emitting region (electron temperature and density, opacity, and optical depth, and the general 3D structure). In short, the emission is a strong function of the viewing angle; the emission strength rapidly decreases as the harmonic ($n$) increases, and the two different magnetic poles can have different magnetic field strength and accretion rates and therefore appear different. A more detailed discussion of the appearance of cyclotron emission can be found in \citet{campbell2008} and \citet{ferrario2015}, and a description of 3D models can be found in \citet{costa2009,silva2013,belloni2021}.

\begin{figure}
    \centering
    \includegraphics[width=\columnwidth]{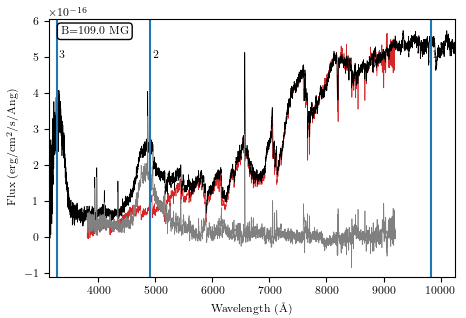}
    \caption{An example (ZTF~J0054+1429) of how we determined the magnetic field strength from the spectra. The black line shows the spectrum, the red line the best-fit red dwarf template, and in grey the difference. The blue vertical lines show the cyclotron harmonic wavelengths for the selected magnetic field strength (the numbers next to the lines indicate which harmonic). We used an interactive version of this figure where the $B$-field strength can be set with a slider to quickly check at which wavelengths the different harmonics should be located.}
    \label{fig:cyclotronexample}
\end{figure}

We analysed the spectra to estimate the magnetic field strength; see, for an example, Fig. \ref{fig:cyclotronexample}. To do this, we first attempted to remove the contribution from the M star in the spectrum. We interpolated the Sloan Digital Sky Survey (SDSS) templates of M dwarfs \citep{bochanski2007}, and fit the best model (M0--M9) using a simple least squares fit. We aimed to fit the continuum redward of \qty{6000}{\angstrom} as well as noticeable TiO bandheads. We resampled the donor star spectrum using the \texttt{specutils}\footnote{\url{https://specutils.readthedocs.io/en/stable/index.html}} package to match the resolution of the observed spectrum. After subtracting the two, we are left with the cyclotron continuum and the white dwarf spectrum (which in most cases contributed only a little). We report the spectral type of the best-fit template in Table~\ref{tab:summary}, rounded to the nearest half spectral type.

We then used an interactive figure to visually inspect the residual spectrum and determine the magnetic field strength. We overplotted the wavelengths of the cyclotron harmonics for an adjustable magnetic field strength; see Equation~\ref{eq:cyclotron} (we assume $\Theta = 90 \degree$). We determined the magnetic field strength by visually matching any broad emission features in the residual spectrum that could be cyclotron emission features, taking into account the wavelengths of the other cyclotron harmonics. Typically, we are able to identify the magnetic field strength if two cyclotron emission features are detected. In some cases, the template subtraction process was not perfect, and a cyclotron peak, typically at longer wavelengths, was erroneously removed. In that case, we repeated the template fitting process, but limited it to longer wavelengths to avoid any strong cyclotron harmonics. 

\subsection{Spectral energy distribution}
In order to estimate the white dwarf temperature, we fit the spectral energy distribution based on archival data, ranging from the far ultraviolet to the far infrared, with white dwarf models \citep{koester2009} and BT-SETTL red dwarf model spectra \citep{allard2003} with $\log g = 5$. The implementation and fitting method is the same as in \citet{vanroestel2021a}, except that we restricted the white dwarf radius between \qtyrange{0.01}{0.014}{\Rsun} corresponding to masses of \qtyrange{0.5}{0.8}{\Msun}, typical masses for white dwarfs in cataclysmic variables. We were unable to measure the white dwarf temperature for all systems because in some cases the cyclotron emission lines strongly affected the SED shape. Furthermore, for some systems, no $u$-band, NUV, or FUV data were available, and the temperature of the white dwarf could not be meaningfully restricted. The temperatures are listed in Table~\ref{tab:summary}. We also used the SED-fit to estimate the spectral type of the M-dwarf, which was typically within one spectral type as determined from the spectra. In cases where no spectral type could be determined from the spectrum, we simply use the spectral type determined from the SED fit.

\subsection{Companion mass, radius, and Roche lobe fill factor}\label{sec:donorcalc}
To determine the mass and radius of the red dwarf and the Roche lobe fill factor, we use a very similar procedure as in \citet{parsons2021}. Briefly, we used the $K$-band magnitude data from 2MASS \citep{Skrutskie2006} along with the parallax measurements provided by \textit{Gaia}, and use the mass-luminosity relationship described in \citep{mann2019} to determine the mass. As in \citet{parsons2021}, we set a lower limit of 5\% on the uncertainty of the mass estimate. We use this mass estimate to infer a radius using a mass-radius relation from \citet{brown2022}, which partially relies on the mass-radius relation from \citet{baraffe2015}. Finally, we calculate the volume of the secondary Roche lobe assuming that the white dwarf has a mass in the range of \qtyrange{0.5}{0.8}{\Msun} and use this to calculate the Roche lobe fill factor by simply dividing the estimated stellar radius by the Roche lobe volume averaged radius ($R_2/R_\mathrm{RL,2}$). The results of the calculations are presented in Table 
\ref{tab:summary}. While uncertainties are numerically propagated in each step, we note that in the calculations of the Roche lobe fill factor, quite a few models are used, which can introduce systematic uncertainties, and we caution against over-interpreting the results.

\section{Results}\label{sec:results}
The results are summarised in Table \ref{tab:summary}. The table lists 22 objects selected based on their ZTF lightcurves and are strong candidates to be low accretion rate magnetic white dwarfs in PCEBs. The first part of the table lists 18 likely wind-accreting magnetic white dwarfs that all show cyclotron humps in their spectra. The second part of the table shows 4 objects that show light curves typical for cyclotron emitting sources, but have much shorter orbital periods. Only one of these objects shows resolved cyclotron humps, while for the other three objects the cyclotron features are likely too broad to be resolved. This is likely caused by a large range in magnetic field strengths, higher temperature, and/or high optical depth, see \citet{campbell2008}. While their light curves stand out in a similar way as the other objects in Table 
\ref{tab:summary}, they are clearly a separate category (see also Sect.~\ref{sec:discussion}).

\begin{figure*}
    \centering
    \includegraphics{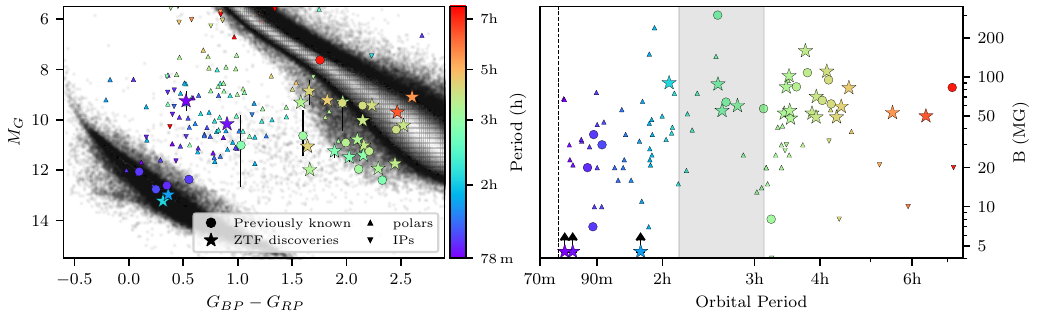}
    \caption{The properties of low accretion rate magnetic white dwarfs in PCEBs. Stars show objects discovered/recovered with ZTF, dots show known systems from \citet{parsons2021}, \citet{hakala2022} and \citet{kawka2021} that we did not recover. The left panel shows the \textit{Gaia} colour-magnitude diagram with the colours of the symbols by the orbital periods. The right panel shows the magnetic field strength versus the orbital period. For additional context, we also show polars (upward pointing triangles) and intermediate polars (downward pointing triangles) from \citet{ferrario2015}. The cataclysmic variable period gap is shown in grey and the CV period minimum as a dotted line \citep[e.g.][]{knigge2011}. The short period ($\lesssim$\qty{2}{hrs}) systems are likely low-state polars or period bounce polars, while the longer period objects are likely pre-polars, detached polars, or low-state polars.}
    \label{fig:HR}
\end{figure*}

Table \ref{tab:summary} also lists the orbital period, companion spectral type, and orbital period of each of the objects. Figure \ref{fig:HR} shows the new systems and previously known systems in the observational colour-magnitude diagram and the magnetic field strength versus the orbital period. 
The orbital periods of the detected systems range from 81 minutes to 6.37 hours. The
The magnetic field strength of the objects is typically \qtyrange{48}{60}{MG}, but the distribution has a long `tail' of stronger $B$-fields. This is again not unusual compared to the sample of known objects. We do note that some of the previously known objects have much lower magnetic fields strengths.
The secondary stars are typically early M dwarfs, although there are a number of objects with late-M or brown dwarf companions, which is again similar to already known objects.

\section{Discussion}\label{sec:discussion}

\subsection{Understanding the light curve shapes}
In order to understand the observed light curve shapes, we must first consider the cyclotron light emitted from a single magnetic pole. Depending on the inclination angle and the latitude of the spot and the optical depth of the harmonic in question, the light curves can have various shapes; approximately sinusoidal (S), pulse-like (P), and `M'-shaped (M); see Appendix \ref{sec:toymodel} for a simple toy-model of a cyclotron emitting spot on a rotating star. Note that the toy-model is overly simplistic, and does not take into account the fact that the emission region can be extended and the optical depth of the emitting region and assumes an optically thin situation. If the spot is optically thick, the radiation pattern changes and the light will be emitted more isotropically (and will appear as pulse-like or sinusoidal-shaped light curves). Note that the optical thickness can be different for different harmonics in the same system. This explains why light curves in different bands ($gri$) can show different shapes (e.g. MQ Dra).    
We also note that the cyclotron emission strength is a strong function of the local magnetic field strength and viewing angle and we refer to \citet{campbell2008} for a detailed discussion of how magnetic field strength, accretion spot temperature, accretion flow density, and viewing angle affect the spectra. 

As shown by \citet{hakala2022}, to understand the various shapes of the light curve, we should also consider extended or multiple emitting regions. Extended emitting regions can smooth the observed signal, and the result of multiple separate emitting regions will result in a mixture of signals, which can also result in an asymmetric light curve shape. However, the light curves of the objects we found can typically be explained well with just one small dominant spot. In addition, even without any cyclotron emission, PCEB binary systems can already show periodic variability in the form of ellipsoidal variation \citep[e.g.][]{gansicke2001}. This is especially visible in the $i$-band light curves; see, for example, the light curve of ZTF~J0054+1429, where no cyclotron peak is present in the $i$-band, but a sinusoidal signal can be seen. 

Despite all these caveats, Fig. \ref{fig:lc_spec1} shows that most of the light curves profiles are approximately consistent with a single spot toy model, although they can differ significantly between filters. Several objects show an `M'-shaped profile in at least one of the bands: 
ZTF~J0548+2638, ZTF~J0646+3958, ZTF~J1218+3409, ZTF~J1531+0204, ZTF~J2000-1435, SDSS~J2229+1853, and ZTF~J2335+3651; 
and also the short period object ZTF~J0112+5827, ZTF~J0343\textminus1655, and ZTF~J1144+3657. The light curve shape is quite unique 
(compared to other types of period variable stars) and is consistent with a high inclination angle and a low spot latitude. There are also a few examples of objects that show a pulse-like light curve profile: ZTF~J0116+4417, ZTF~J0516+0354, ZTF~J0542+0518, and ZTF~J0900+6216. These objects are likely viewed at a high inclination angle and the magnetic spots are located close to the north or south pole. Finally, there are some more sinusoidal light curves: ZTF~J0054+1429, ZTF~J0257+3328, ZTF~J1239+7041, MQ~Dra, and ZTF~J1723+3427 and short period system ZTF~J0146+4914. These objects probably have spots close to the rotation axis and are viewed at low to intermediate inclination angles. 

Although most light curves seem to be dominated by a single spot, a few light curves show hints of two spots rotating in and out of view the previously known system WX~LMi is a great example (Fig. \ref{fig:lc_spec1}). New discovered systems that likely show two spots are ZTF~J1531+0204, ZTF~J2000-1435, and possibly ZTF~J2335+3651. These light curves show a `M'-shaped or pulse-shaped light curve, as well as an additional lower-amplitude peak offset by half a phase. This suggests that in all cases, one spot is clearly dominant and the second spot at best only contributes slightly to the overall luminosity. This finding is consistent with what is seen using phase-resolved spectra by \citet{parsons2021} and \citet{hakala2022}.  

\begin{figure*}
    \centering
    \includegraphics{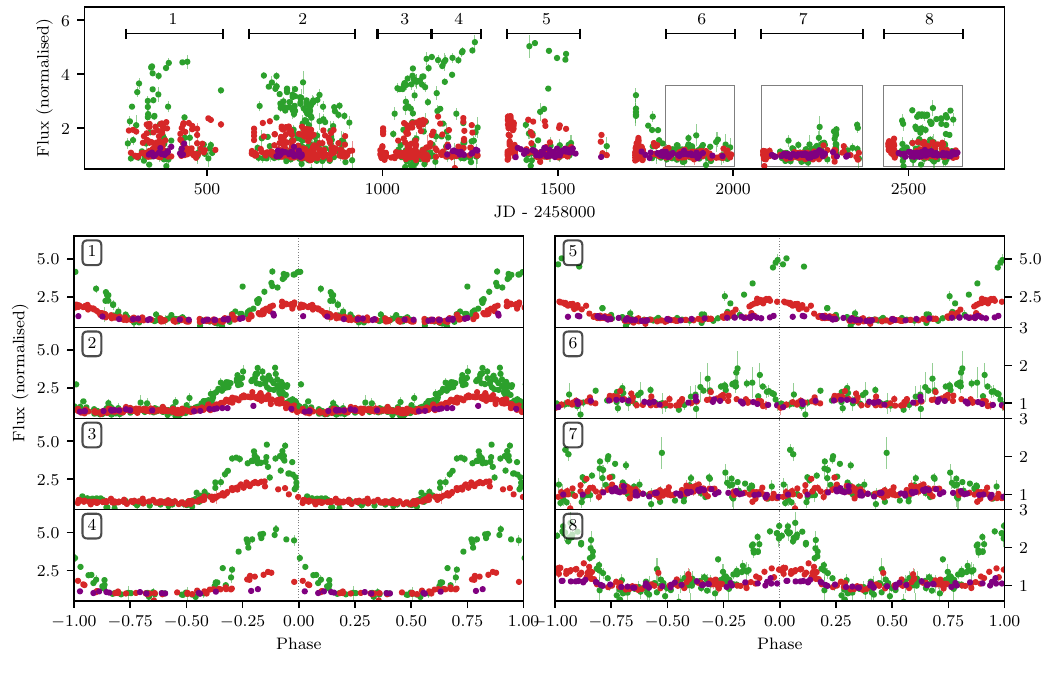}
    \caption{The ZTF light curve of ZTF~J0116+4417. The top panel shows the full light curve and clearly shows that the amplitude of variability changes over a timescale of years. In the bottom panels, we show the folded light curves per part of the data as indicated in the top panel. The y-scale is the same for each panel, except for the bottom-right panel, which shows a more zoomed-in view. The bottom panels show that both the amplitude and the peak-phase change slowly over time. For example, the difference in peak-phase between year 1 and year 2 is almost 0.2 in phase, and the amplitude change between panel 3 and 4 is almost a factor of 2.}
    \label{fig:lc_J0116}
\end{figure*}

\subsection{Long timescale changes in the light curve shape}\label{sec:modulation}
Some systems show periodic signal that slowly change over years, for example ZTF~J0116+4417, SDSS J1514+0744, and MQ~Dra.
ZTF~J0116+4417 shows by far the largest changes and we examine this system in more detail; see Fig. \ref{fig:lc_J0116}. The first observation is that the amplitude of the cyclotron emission is modulated on a long time scale, $\approx$2.4 years according to the Lomb-Scargle periodogram. This is especially visible in the $g$-band. In addition to the changes in the amplitude, we also see that the cyclotron emission peak shifts in phase, by up to 0.2 between year 1 and year 2 (see the lower panels of Fig. \ref{fig:lc_J0116}). 

 We consider two possible physical mechanisms for these long time-scale changes. The first possibility is that the magnetic field strength or geometry of the binary changes over time. This would change the geometry of the accretion flow and therefore also the shape of the emission as a function of wavelength and/or orbital phase. An example of a polar that slowly changes its geometry is DP Leo \citep{beuermann2014}. In that case, the magnetic axis of the white dwarf is librating with a period of $\approx$\qty{60}{yrs} which is observed as a slow change in the longitude of the magnetic accreting spot. It is possible that something similar is happening in the case of ZTF~J0116+4417, but we note that the timescale is shorter than the theory predicts \citep{king1990,king1991,wickramasinghe1991}. 

 Another possibility is that the mass transfer rate from the secondary star changes, e.g. because of starspots. This could change both the temperature of the accretion region and the particle density.  Both explanations are possible and long timescale spectroscopic monitoring of this system would be able to shed more light on what is driving the observed long-timescale changes.

\subsection{Comparison with known systems, selection biases and completeness}\label{sec:completeness}
In this section, we briefly discuss the biases and completeness of (ZTF) multicolour light curves as a tool to discover low accretion rate magnetic white dwarfs that emit cyclotron radiation. 
The objects presented in this work were found serendipitously as anomalous periodic variable stars in ZTF data. This means that a strong cyclotron peak must be present in the $g$ or $r$ bands. This sets an upper limit to magnetic field of $B \lesssim \qty{250}{MG}$, as the $n$=1 harmonic moves out of the optical bandpass and into the UV for stronger fields (see Equation \ref{eq:cyclotron} and \citealt[]{hakala2022}). For $B \lesssim \qty{90}{MG}$ (or weaker fields), the 2nd and 3rd harmonics (or higher order harmonics) are present in the $g$ and $r$ bands, which makes these objects easier to identify. 
The lower limit is not determined by the cyclotron wavelengths but by the (relative) cyclotron emission strength. Based on the magnetic field strength of the recovered systems, we estimate that the detection limit is $\approx \qty{40}{MG}$ (when only $n>4$ harmonics are present in the $g$ and $r$ bands).
The second requirement is that the cyclotron emission is highly variable as a function of the orbital phase, discussed in the previous section. In addition, even if cyclotron emission is responsible for the variability, the signature is not always unique enough to recognise it as such. The `M'-shaped and pulse-like light curves are quite unique; however, sinusoidal variability can be the result of ellipsoidal variability or starspots. In summary, the light curve properties of the known sample and new sample are quite different, mostly in variability amplitude, because of the methods used to identify the new and previously known objects.

This suggests that both samples (the previously known and the ZTF sample) are highly incomplete, and we only find a small fraction of systems, even if there is cyclotron emission in the $g$ and $r$ bands.
However, if we compare the measured properties (magnetic field strength, orbital periods, and white dwarf temperature), we do not observe any significant differences between the new objects and previously known objects (see Sect. \ref{sec:results} and Fig. \ref{fig:HR}). This is likely due to biases in both samples for a specific geometry of systems. The largest and most unique photometric variability is seen for high inclinations and low spot latitudes. This bias is reversed for a spectroscopically selected sample; such a sample would be biased towards objects that show cyclotron emission over the entire rotation phase (low inclinations and high spot latitudes; see Sect. \ref{sec:toymodel}). That the magnetic field strengths and orbital periods are mostly similar for the new and previously known samples suggests that the selection bias for these parameters is not as strong as the bias for the geometry of the systems. 

Besides the geometry of the systems, the mass-accretion rate is another possible explanation for the observed amplitude difference in the light curves (as we suggested in the previous section to explain the long term variability in ZTF J0116+4417). \citet{parsons2021} has estimated the mass-accretion rate in a few systems. Unfortunately, we lack the high-quality, time-resolved spectra to do the same for the newly discovered objects. However, if we compare the ZTF light curves of systems with measured mass-accretion rates from Table~4 in \citep{parsons2021}, we do not observe an obvious correlation between the mass-accretion rate and the variability amplitude. For example, MQ Dra and SDSS~J2229+1853 are previously known objects that we recovered because of their large amplitude of variability, but the former has one of the lowest estimated mass-transfer rates (\qty{0.6e-13}{\Msun yr^{-1}}) while the latter has one of the highest mass-transfer rates (\qty{6.5e-13}{\Msun yr^{-1}}). This does not exclude that the mass-accretion rate correlates with the observed amplitude; but it cannot be the dominant (or only) factor that determines the amplitude. 

Finally, we try to estimate the completeness of our search of the ZTF data.
The ZTF light curves of the previously known systems that we did not recover are shown in Fig. \ref{fig:lc_known} and \ref{fig:lc_known_polars}. These were discovered primarily spectroscopically (mostly by the SDSS survey), or because they are eclipsing and the eclipse shape shows a subtle but significant cyclotron signature \citep[e.g.][]{parsons2015}.  Most of the light curves show only subtle variability or no visible periodic variability at all, and therefore most have not been identified as cyclotron-emitting PCEBs by their ZTF light curve. 

Out of the known objects we recovered 4 known PrePs/LARPs. In addition, we also identified the two eclipsing systems (SDSS~J0303+0054 and SDSS~J1206+5100) as cyclotron emitting systems, but this was greatly aided by the fact that they are eclipsing. If we limit ourselves to the non-eclipsing systems, the recovery rate is 4 out of 13 previously known objects. Since the spectroscopic searches (the main source of previously identified objects) are also incomplete, this puts an upper limit on the completeness of $\lesssim$ 0.3 for cyclotron emitting, low accretion rate white dwarfs that are in the ZTF footprint. Because the spectroscopic sample we use as a reference is also incomplete, this completeness estimate is an upper limit.

\subsection{Evolutionary status}

\begin{figure}
    \centering
    \includegraphics{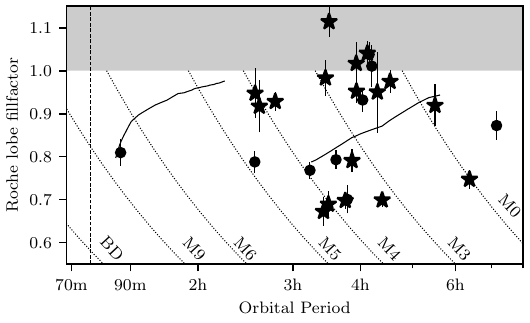}
    \caption{The best estimate of Roche lobe fill factor ($R_{2} / R_{L,2}$) 
    for the donor star in each system, adapted from \citet{parsons2021}. Dots show previously known system, stars show the ZTF detected objects. Note that a few assumption have been made in order to calculate the Roche lobe fill factor, see Section \ref{sec:donorcalc}.
    The dashed lines show the tracks for different companion stars if only the orbital period changes. Solid lines shows how much a cataclysmic variable would shrink if it were to detach \citep{knigge2011}. The vertical dashed line shows the cataclysmic variable period minimum.
    The figure shows that some systems are clearly filling their Roche lobe and are likely low-state polars, while other systems are quite far from filling their Roche lobe and could be pre-polars. Note that systems that are filling their Roche-lobe (fill factor=1) can result in the mass-estimate and therefore fillfactor estimate, to be overestimated.
    }
    \label{fig:RLFF}
\end{figure}


As we discussed earlier in the Introduction (Sect. \ref{sec:intro}), there are multiple evolutionary phases of magnetic white dwarfs in PCEBs where the accretion rate is low: pre-polars \citep{schwope2009}, low-state polars (e.g. EF~Eri), detached polars \citep{schreiber2021}, or (detached) magnetic period bouncers \citep{breedt2012,kawka2021,schreiber2023}. See also \citealt{parsons2021} for an extended discussion on the evolutionary status of the previously known objects. In this section, we briefly discuss the evolutionary status of each of the objects we identified.

First, we consider the white dwarf temperatures; they are all in the range of \qty{6500}{} to \qty{12000}{K}, similar to previously known objects. The fact that we are only finding low temperature white dwarfs is consistent with observations of magnetic single white dwarfs by \citet{landstreet2019,bagnulo2020}. They showed that the incidence rate of magnetic fields is much higher in old and cold compared to young and hot white dwarfs. If we compare temperature of the white dwarfs to the white dwarf temperature in non-magnetic CVs in the period range of 3-7 hour, we see that the white dwarf temperatures in CVs are much higher (due to compressional heating), see e.g. \citet{townsley2009,pala2017}. This makes it unlikely that the ZTF discovered systems have shown higher mass-transfer rates (Roche lobe overflow) in the recent past, otherwise the white dwarf would have been much hotter. 

The most informative binary parameter is the Roche lobe fill-factor, shown in Fig. \ref{fig:RLFF}.
 Note that we did have to make some assumptions to do so, so we caution the reader not to over-interpret the measured values (see Sect. \ref{sec:methods}). The figure shows that some systems clearly fill their Roche lobe and have Roche lobe fill factor values of 1 or more. These are possibly polars that are in a low-state, and mass-transfer from the donor has only temporarily stopped, like for example EF~Eri. There are a few systems that are mildly detached systems (0.8--0.9) that could be detached magnetic cataclysmic variables and would fit the formation scenario proposed by \citet{schreiber2021}. However, there are also a few systems that have Roche lobe fill factors lower than $\lesssim 0.8$. These systems that are far from filling their Roche lobe, and if the measured Roche lobe fill factors are correct, are only consistent with being pre-polars; systems that have never experienced Roche lobe overflow accretion and emerged from the common envelope as we observe them today. In particular, the two longest period systems (with periods greater than 6 hours) are difficult to reconcile as detached CVs and are most likely pre-polars.

We now focus our attention on the short period systems (bottom of Fig.~\ref{tab:summary} and Fig.~\ref{fig:lc_spec_polars}). 
If we inspect Fig.~\ref{fig:HR}, we notice that the two shorter period objects (ZTF~J0112+5827 and ZTF~J0343-1655, both with $P_\mathrm{orb}$=\qty{81}{m}) are located above the white dwarf track where most of the polars are also found. This, together with the fact that their orbital period is close to the CV period minimum and is similar to the majority of polars, strongly suggests that these two objects are polars that are currently in an extended low state. For example, known polars such as EF~Eri or IL~Leo also show year-long low states where they show a similar strictly periodic light curve with a similar profile.

The two longer period objects (ZTF~J0146+4914, $P_\mathrm{orb}$=\qty{123}{m}; and ZTF~J1144+3657, $P_\mathrm{orb}$=\qty{109}{m}) are right on the white dwarf track and also show little or no infrared excess in the SED. This strongly suggests that the donor is cold and does not contribute to the optical brightness and is likely a brown dwarf. This leaves two possibilities; either these systems are simply pre-polars that were born with a brown dwarf companion, or they are magnetic period bouncers. Magnetic period bouncers are the end state of magnetic CV evolution where the donor lost most of its mass and has become a brown dwarf \citep{breedt2012,kawka2021,schreiber2023}. In this phase, the orbital period of the system is slowly increasing, and the Roche lobe mass-transfer rate is very low, similar to the non-magnetic period-bouncers (e.g. \citealt{pala2018,inight2023}). The white dwarf temperature for period bouncers is $\approx$\qty{9000}{K}, which is similar to the non-magnetic systems \citep{pala2017}. Finally, the fact that we are only seeing resolved cyclotron humps for the longest period system is consistent with this picture; only for very long orbital period period-bouncers the accretion rate is low enough (and similar to a wind-accretion rate, $\dot{\mathrm{M}} \lesssim10^{-13}$ \unit{M_\odot yr^{-1}}) and optical depth low enough for individual cyclotron humps to show up in the optical spectrum. We note that the known candidate magnetic period bounce candidates have periods of $\approx$\qty{90}{m} (Fig. \ref{fig:lc_known_polars}) and show almost no variability. Finally, we also see that for the two objects that have an estimated Roche-lobe fill factor, the value is about $\approx$ 0.8, suggesting that these systems are detached.
Precise and accurate measurements of the binary parameters (especially the Roche-lobe fill-factor) and the magnetic field strength (possibly by detecting Zeeman splitting) for these systems are needed to better understand their evolutionary status.

In summary, the evolutionary history of the new discovered ZTF objects is uncertain, and we found at least a few objects consistent with each of the various evolutionary scenarios. Precise and accurate white dwarf mass and donor Roche lobe fill factor measurements are needed to better understand the population of these objects and understand the history of this type of objects (e.g., using eclipsing systems that allow for model independent parameter measurements).

\section{Prospects and future work}\label{sec:futurework}
In this work, we show that multi-colour light curves can be used to identify low accretion rate, cyclotron emitting PCEB binaries. This method can also be applied to other deep multi-colour photometric survey data such as ATLAS \citep{tonry2018}, BlackGEM \citep{groot2019}, \textit{Gaia} \citep{eyer2023}, and the Rubin Observatory \citep{hambleton2023}. The broader wavelength coverage offered by a broader filter set of LSST ($ugrizy$) and BlackGEM ($ugriz$) are useful to observe the variability of multiple harmonics. 
The addition of the $u$-band would be especially useful because contribution by the secondary star are minimal (e.g. spot variability, ellipsoidal variability) while the cyclotron emission can be very strong (e.g. ZTF~J0054+1429 in Fig.~\ref{fig:lc_spec1}). 
This obviously extends into the UV, and the type of systems discussed in the paper will also be detected as unusual periodic variable stars in any UV time-domain survey, e.g. UVEX \citep{kulkarni2021}.
In addition, a more targeted and automated method can be used to identify light curve shapes consistent with cyclotron emitting PCEBs. This can be done by fitting a model to light curves, but more interesting (and effective) would be the use of anomaly detection methods \citep[e.g.][]{rebbapragada2009,chan2022}.

As discussed in Section \ref{sec:completeness}, for more than half of the known objects the ZTF light curve does not show enough variability to identify a system as a cyclotron-emitting PCEB. 
However, as demonstrated in this paper, a (low-resolution) spectrum is enough to identify cyclotron emission lines. Large samples of spectra by e.g. \textit{Gaia} \citep{deangeli2023}, SDSS~V \citep{kollmeier2017}, DESI \citep{desicollaboration2016}, WEAVE \citep{dalton2016}, and 4MOST \citep{dejong2019} would likely yield more cyclotron-emitting PCEB binaries. However, these surveys will also not be complete since the systems do not show cyclotron emission features at all orbital phases. In addition, the cyclotron emission strength can be variable over timescales of years. We therefore conclude that, in order to get a complete sample of wind-accreting white dwarfs in PCEBs, both photometric and spectroscopic survey data should be searched. A possible alternative method to find more of these systems is by using X-ray data, for example by SRG/eROSITA \citep{Rodriguez2024}. The archetypal LARP, EF~Eri, was shown to have an X-ray luminosity lower than that of the population of polars \citep{schwope2007}. Since SRG/eROSITA will be the deepest all-sky X-ray survey (and, for example, capable of detecting EF~Eri), many more such systems are likely to be found \citep[e.g.][]{schwope2024}.

In addition to getting a more complete sample of systems, known systems should also be studied in more detail as was done by \citet{parsons2021} and \citet{hakala2022}. As they showed, precision studies can be used to determine the magnetic field geometry, but also the Roche lobe fill factor of the secondary star, which is important to infer the evolutionary history of the system and to test recent evolutionary model predictions \citep{schreiber2022,schreiber2023}. Eclipsing systems will be especially useful for precisely measuring the binary parameters and the spot geometry, and in addition, a sample of eclipsing systems could have a better understood bias in part because lower strength magnetic fields can be detected \citep[e.g.][]{parsons2013a,brown2022}. 


\section{Summary and conclusion}\label{sec:conclusion}
We summarise our paper as follows:

\begin{itemize}
    \item  We present a search for cyclotron emitting PCEBs using the Zwicky Transient Facility (ZTF) light curves. We discovered 14 new pre-polars or low accretion rate polar, doubling the currently known sample of these objects. In addition, we recovered three low-state polars that might be magnetic period bouncers, and identified one new candidate object. To accomplish this, we used the periodic variability detected in the ZTF light curves. Cyclotron `humps' were detected in low-resolution follow-up spectra for all but three objects (short period systems), allowing the estimation of the magnetic field strength. For the three systems where we could not detect cyclotron features in the spectra (short period systems), the light curve was distinctive enough to confidently identify them as magnetic wind-accreting systems.
    \item Using the light curves, spectra, parallax and archival photometry data we measured the orbital period, magnetic field strength, white dwarf temperature, and donor mass, spectral type, radius, and Roche lobe fill factor. 
    Despite the different selection methods compared to the known sample of objects (light curve versus spectroscopic), the physical properties (orbital period, magnetic field strength, white dwarf temperature) are similar between the two samples. We conclude that the differences in the geometry (inclination, spot-latitude), partially the magnetic field strength, and possibly the wind-accretion rate are the cause of the differences in appearance.
    \item This paper presents the first time that light curve shapes have been used to identify cyclotron emitting sources. The discovery and recovery of known sources demonstrates that both colour differences in phase-folded light curves and the light curve shapes can be used to identify cyclotron-emitting white dwarf binaries. The selection method is relatively clean with spotted stars the most common type of false positive that can be rejected on close inspection of the multi-band light curves. The method is far form complete,  $\lesssim$30\%, and seems mostly senstive to high magnetic field systems. However, it does have different biases compared to spectroscopic searches and both methods complement each other. Both future photometric, spectroscopic, and possibly high energy surveys will be able to find more of these cyclotron emitting systems, getting us a more complete picture of the population of these objects.
    \item At least some wind-accreting systems can show large changes in amplitude and phase in their phase-folded light curve over timescales of  $\approx$years. This could be due to changes in the wind-accretion rate, but another possibility is libration of the white dwarf magnetic field axis.
    \item The evolutionary state of these systems is uncertain. Some systems are consistent with being low-state polars (LARPs), while another part would be consistent with being pre-polars (PrePs) or detached polars. Four systems that have short orbital periods have a brown dwarf as companion, and these could be period bounce systems polars. It is unclear if the donor in these systems are filling their Roche lobe or have permanently detached and feature wind-accration.   
\end{itemize}

\begin{acknowledgements}
We thank the referee for the careful review of our manuscript.

We thank Tom Marsh, Steven Parsons, and Axel Schwope for useful discussions.

This publication is part of the project "The life and death of white dwarf binary stars" (with project number VI.Veni.212.201) of the research programme NWO Talent Programme Veni Science domain 2021 which is financed by the Dutch Research Council (NWO).
ACR acknowledges support from an NSF Graduate Fellowship. 
This research was supported in part by the National Science Foundation under Grant No. NSF PHY-1748958.

Based on observations obtained with the Samuel Oschin Telescope 48-inch and the 60-inch Telescope at the Palomar Observatory as part of the Zwicky Transient Facility project. ZTF is supported by the National Science Foundation under Grants No. AST-1440341 and AST-2034437 and a collaboration including current partners Caltech, IPAC, the Oskar Klein Center at Stockholm University, the University of Maryland, University of California, Berkeley, the University of Wisconsin at Milwaukee, University of Warwick, Ruhr University, Cornell University, Northwestern University and Drexel University. Operations are conducted by COO, IPAC, and UW.

The SED Machine is based upon work supported by the National Science Foundation under Grant No. 1106171.
This work used the Apache Point Observatory 3.5m telescope, which is owned and operated by the Astrophysical Research Consortium.
This work used the Palomar Observatory 5.0m Hale telescope, which is owned and operated by the Caltech Optical Observatories.
The data presented herein were obtained at the W.M. Keck Observatory, which is operated as a scientific partnership among the California Institute of Technology, the University of California and the National Aeronautics and Space Administration. The Observatory was made possible by the generous financial support of the W.M. Keck Foundation. 
Based on observations made with the Isaac Newton Telescope (INT) operated on the island of La Palma by the Isaac Newton Group in the Spanish Observatorio del Roque de los Muchachos of the Instituto de Astrofisica de Canarias. 

This research has made use of the SIMBAD database, operated at CDS, Strasbourg, France. 

\end{acknowledgements}

%
%

\bibliographystyle{aa}
\bibliography{zotero_test.bib} 

\begin{appendix}

\section{A toy light curve model}\label{sec:toymodel}
 We consider a simple toy model to help us understand how the geometry determines the basic shape of the light curve of a rotating white dwarf that emits cyclotron radiation. A similar model but for synchrotron radiation instead of cyclotron radiation is presented in \citet{potter2018}. We also note that more sophisticated models exist; see, for example, \citet{schwope1990}, \citet{schwope1993}, \citet{gansicke2001},  \citet{schwope2003}, \citet{kolbin2022}, \citet{hakala2022} and also the \textsc{cyclops} code by \citet{costa2009,belloni2021}. These implement the physics of magnetic accretion in a much more advanced and refined way, which allows for a quantitative comparison between the model and data. We emphasise that the goal of the toy model is to understand the basic differences and properties of light curve shapes qualitatively, for which a simple toy model is sufficient. 

\begin{figure}[h!]
    \centering
    \includegraphics[width=0.48\textwidth]{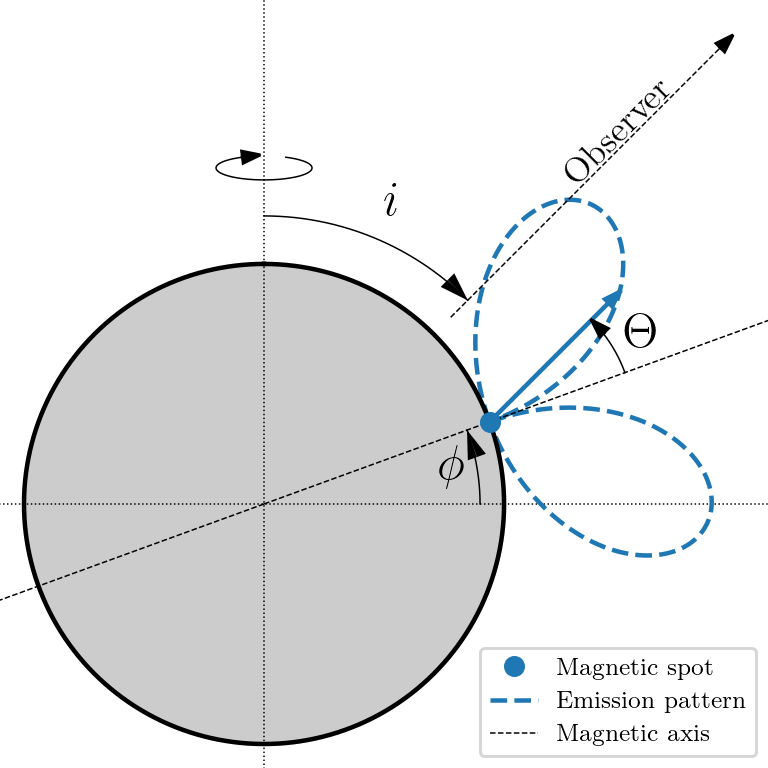}
    \caption{A toy-model of a rotating star with an emitting spot with a angular dependent emission pattern. The diagram shows a 2D crosssection for the phase when the spot is closest to the observer. The y-axis is the rotation axis of the sphere.
    The inclination of the system ($i$) is defined as the angle between the rotation axis and observer line-of-sight. The spot-latitude ($\phi$) is defined as angular distance from the equator. The emission pattern indicates the emission intensity depending on the viewing angle ($\Theta$, the 3D angle between the magnetic axis and observer line-of-sight).}
    \label{fig:toy_diagram}
\end{figure}

The toy model (see Fig. \ref{fig:toy_diagram}) consists of an opaque rotating sphere viewed at an inclination angle $i$ (0\degree\ to 90\degree, the angle with the rotation axis and the observer). An emitting point source spot is located on the surface of the sphere with latitude $\phi$ (\textminus90\degree\ to 90\degree). The emission strength of the spot depends on the viewing angle: $I\propto \sin(\Theta) \cos(\Theta)$, with $\Theta$ the angle between the observer and normal vector of the surface at the location of the spot. The $\sin(\Theta)$ term is typical for cyclotron radiation which preferentially radiates perpendicular to the magnetic field lines (here we assume that the magnetic field lines are normal to the surface of the sphere, see also \citealt{campbell2008}). The $\cos(\Theta)$ term corrects for the geometric projection of the spot. 
We note that this is only valid for the optically thin case, typically the lower order harmonics.
 
The combination of these terms has the consequence that the highest intensity occurs at a viewing angle $\Theta$=45\degree, and no photons are detected when viewed from directly above $\Theta$=0\degree. The blue dashed lines in Fig. \ref{fig:toy_diagram} show this emission pattern. For comparison to an observed emission pattern of cyclotron emission see Fig. 5 in \citet{schwope1990}.

\begin{figure*}[h!]
    \centering
    \includegraphics{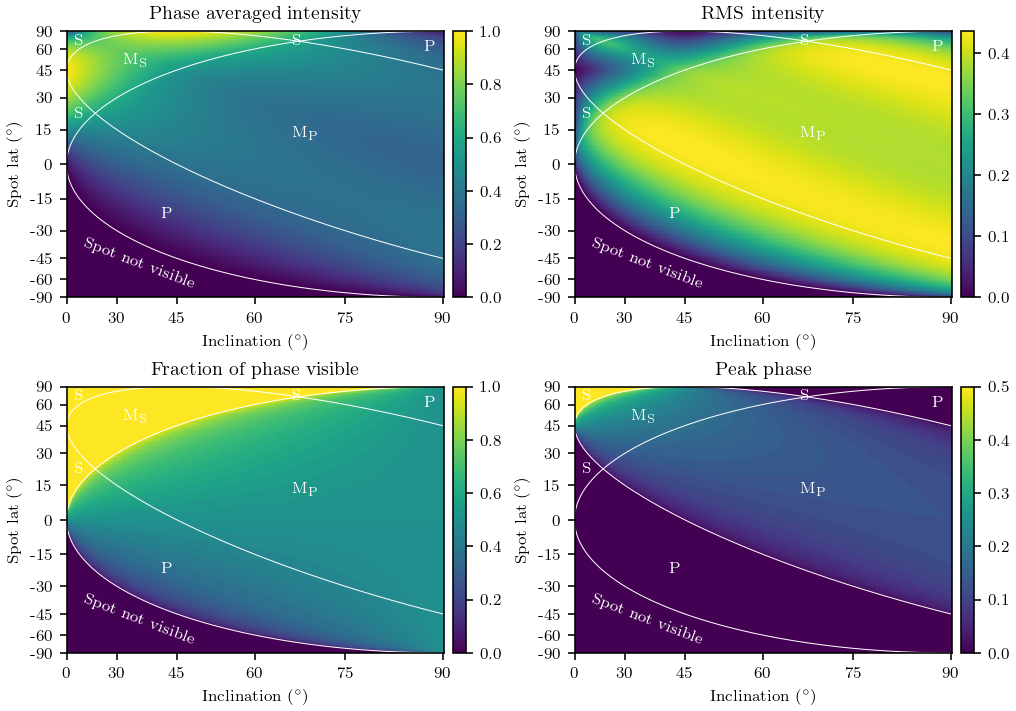}
    \caption{An overview of the light curve properties of the toy-model as function of the inclination angle ($i$) and spot latitude ($\phi$). The figure shows that the light curve properties (average, amplitude, and shape) vary greatly depending on the geometry of the system. The white lines delineate the different light curve shapes. Examples of the different light curve shapes are shown in Fig. \ref{fig:toy_lcs}}
    \label{fig:toy_overview}
\end{figure*}

\begin{figure*}[h!]
    \centering
    \includegraphics{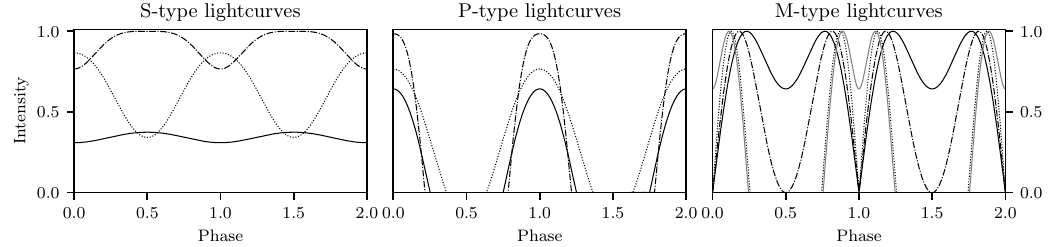}
    \caption{Model light curves for the different light curve shapes. The left panel shows approximately sinusoidal light curves from the `S'-labelled regions in Fig. \ref{fig:toy_overview}. They can have a range of amplitudes, average values, and shapes. The examples shown are for $i$=10\degree\ and $\phi$=89\degree, 55\degree, 20\degree\ (full, dashdotted, dotted). 
    The middle panel shows pulse-like light curves, labelled `P' in Fig. \ref{fig:toy_overview}. They vary both in amplitude and width, but during atleast some of the phase the intensity is 0. 
    The examples shown are for [$i,\phi$] = [20\degree,~0\degree],[60\degree,~$-20$\degree], [80\degree,~75\degree]; (full, dashdotted, dotted). The right panel shows `M'-shaped light curves, labelled `M' in Fig. \ref{fig:toy_overview}. The examples shown are for with $i$=35\degree, 45\degree, 80\degree\ and $\phi$=90\degree\textminus$i$ (full, dashed-dotted, dotted). The grey line shows a model $i$=80\degree\ and $\phi$=$-10$\degree.}
    \label{fig:toy_lcs}
\end{figure*}

With this simple toy model, we can produce various light curves as functions of rotation phase. Figure \ref{fig:toy_overview} shows various characteristics of the light curve as a function of the angle of inclination ($i$) and the latitude of the spot ($\phi$). We divide the light curves into three types: pulse-like profiles (P), approximately sinusoidal profiles (S), and `M'-shaped profiles (M); see Fig. \ref{fig:toy_lcs} for examples. 
\begin{itemize}
\item P-type light curves show a pulse-like shape where for a significant part of the light curve no cyclotron light is detected. These occur for orientations where the spot is not visible for some part of the rotation of the star, and when the spot is visible, the viewing angle ($\Theta$) is always larger than 45\degree. 
\item S-type light curves appear for objects where the spot is visible in all orbital phases, but the viewing angle is always larger than 45\degree. 
\item M-type light curves occur when the viewing angle ($\Theta$) is less than 45\degree\ at some phase of rotation. We subdivide the different light curves in $\mathrm{M_P}$ and $\mathrm{M_S}$ subtypes. The $\mathrm{M_P}$ light curves show a narrow double-peaked profile with M-shaped peaks and show zero emission at other phases. The $\mathrm{M_S}$ light curves also show a double-peaked profile, but they are continuous and never show a zero-flux phase.
\end{itemize}

In summary, this simple toy model shows that the observed light curve strongly depends on the inclination and spot latitude. This includes not only the light curve shape, but also the amplitude and rotation-averaged emission strength of the spot. Therefore, the geometry of the system strongly determines if such systems can be detected and recognised by their light curves.

\onecolumn

\clearpage

\twocolumn

\section{Figures and Tables}
The cyclotron emitting objects discovered or recovered in ZTF are listed in Table~\ref{tab:summary} and their light curves and spectra are shown in Figure~\ref{fig:lc_spec1} and \ref{fig:lc_spec_polars}.

Figure \ref{fig:lc_known} shows the ZTF light curves of known wind-accreting magnetic white dwarfs in PCEBs as listed in \citet{parsons2021} and \citet{hakala2022}. Figure \ref{fig:lc_known_polars} shows the light curves and spectra of polars we identified as candidate cyclotron emitting systems and we followed up with a spectrum.
Figure \ref{fig:lc_known2} shows two known wind-accreting systems that switched to a higher accreting rate and should therefore be classified as polars. 


\begin{landscape}
\begin{table}
\tiny
\renewcommand{\arraystretch}{1.2}
  \begin{tabular}{lllllllllllll}
    \hline
ID &  $G$ & distance  & P$_\mathrm{Orb}$ & WD temp &  Sp. type & $M_2$ & RLFF & {\bf B}-field & spectra & Peaks & Harmonics  \\
   & \unit{AB mag} & \unit{pc} & \unit{h} & \unit{kK} & & \unit{\Msun} & & \unit{MG} & & \unit{\angstrom} & & \\
\hline
\\
\multicolumn{9}{l}{Candidate wind-accreting systems} \\
\hline
ZTF~J005400.33+142931.4 & 17.40 & \phantom{0}$393^{+18}_{-15}$ & 4.12 & $11.1_{-0.3}^{+0.3}$ & M3.5 & $0.461\pm0.023$ & $1.04\pm0.03$\, & $110.5 \pm 0.5$ & SEDM, LRIS, DIS & 4856 ($g$), 3230 &  2, 3  \\
ZTF~J011655.02+441745.2 & 18.70 & \phantom{0}$923^{+275}_{-171}$ & 4.30 & $12.0_{-1.5}^{+1.8}$ & M2.0 & $0.41\phantom{0}\pm0.08$ & $0.95\pm0.09$ & 49 & LRIS & 7315? ($i$), 5486 ($g,r$), 4389 ($g$), 3657 & 3, 4, 5, 6 \\ 
ZTF~J025740.49+332808.3 & 18.91 & \phantom{0}$243^{+16}_{-13}$ & 3.49 & $8.1_{-0.5}^{+0.8}$ & M6.0 & $0.166\pm0.013$& $0.69\pm0.03$ & 204/102 & SEDM, DIS & 5260 ($g$) & 1/2 \\
ZTF~J051623.71+035400.3 & 18.68 & \phantom{0}$223^{+13}_{-11}$ & 3.42 & $6.6_{-0.4}^{+0.3}$ & M4.0 & $0.155\pm0.013$& $0.67\pm0.03$\, & $52.5 \pm 1$ & Kast & 6806 ($r$), 5104 ($g$) & 3, 4 \\
ZTF~J054247.38+051856.1 & 19.15 & \phantom{0}$342^{+32}_{-30}$ & 2.55 & $7.0_{-0.8}^{+0.8}$ & M4.0 & $0.205\pm0.025$& $0.95\pm0.06$\, & 88  & DBSP & 6155 ($r$), 4103 ($g$) & 2?, 3  \\
ZTF~J054800.74+263821.4 & 16.66 & \phantom{0}$193^{+3}_{-3}$ & 3.93 & $7.9_{-1.1}^{+1.1}$ & M4.5 & $0.371\pm0.019$& $0.95\pm0.03$\, & 70.5 $\pm$ 0.5  & DBSP  & 7574 ($i$), 5049 ($g$), 3787 & 2?, 3, 4 \\
ZTF~J064618.65+395837.9 & 20.03 & $1410^{+1010}_{-545}$ & 3.51 & - & M4.0 & - & - & 48 $\pm$ 0.5 & LRIS & 7595 ($i$), 5696 ($r$), 4557 ($g$) & 3, 4, 5 \\
ZTF~J090018.54+621657.4 & 16.91 & \phantom{0}$277^{+4}_{-4}$  & 6.37 & $8.4_{-0.4}^{+0.5}$ & M3.5 & $0.428\pm0.022$& $0.75\pm0.02$\, & 49.5 $\pm$ 0.1 & SEDM &  7226? ($i$), 5420 ($g$), 4336 ($g$)&  3, 4, 5 \\
\textit{WX LMi}$^1$ & 16.34 & \phantom{00}$96.8^{+0.5}_{-0.5}$ & 2.78 & $8.3_{-0.6}^{+0.7}$ & M5.0 & $0.222\pm0.011$ & $0.93\pm0.02$ & 69.9 & DBSP, SDSS & 5100 ($g$) &  3  & \\ 
ZTF~J121804.90+340921.9 & 19.04 & \phantom{0}$364^{+52}_{-36}$ & 2.60 & $8.5_{-0.7}^{+0.7}$   & M3.5 & $0.20\pm0.027$& $0.92\pm0.06$\, & $55\pm0.5$  & SEDM & 6490 ($r$), 4868 ($g$), 3894 & 3, 4, 5 \\
ZTF~J123903.50+704136.9 & 17.89 & \phantom{0}$169^{+2}_{-2}$ & 3.75 &  $6.7_{-0.6}^{+0.7}$  & M5.0 & $0.186\pm0.009$ & $0.70\pm0.02$\, &  $159\pm0.5$  & DBSP & 6730 ($r$)&  1 \\
ZTF~J153134.34+020458.8 & 18.46 & \phantom{0}$307^{+16}_{-16}$ & 3.86 &  $9.1_{-0.7}^{+1.0}$  & M5.5 & $0.251\pm0.017$& $0.79\pm0.03$\, & $52.5\pm0.5$  & SEDM &  6860 ($r$), 5150 ($g$) &  3, 4 \\
\textit{MQ Dra}$^2$ & 17.34 & \phantom{0}$181^{+2}_{-2}$ & 4.39 & $7.8_{-0.5}^{+0.5}$ & M5.0 & $0.231\pm0.011$& $0.70\pm0.02$\, & $59.2 \pm 0.2$\,  & SEDM & 5970 ($r$), 4477 ($g$) & 3, 4 \\
ZTF~J172333.86+342730.8 & 18.08 & \phantom{0}$286^{+5}_{-5}$ & 3.50 &  $8.8_{-1.4}^{+2.0}$ & M7.0 & $0.428\pm0.024$&  $1.11\pm0.03$\, & 174/87/58  & SEDM &  6155 ($r$) & 1/2/3?\\
ZTF~J192001.0+772415.1 & 18.67 & \phantom{0}$821^{+119}_{-98}$ & 5.50 & $8.4_{-1.5}^{+1.4}$ & M4.0 & $0.52\pm0.05$ & $0.92\pm0.05$ & 53? & DBSP & 6730? ($r$), 5050? ($g$) & 3?,4?  \\ 
ZTF~J200034.84-143534.4 & 18.58 & \phantom{0}$357^{+28}_{-26}$ & 3.44 & $9.4_{-0.9}^{+1.3}$ & M5.5 & $0.333\pm0.027$& $0.98\pm0.04$\, & $83.5\pm0.5$ & IDS & 6405 ($r$), 4270 ($g$) & 2,3 \\
\textit{SDSS J222918.97+185340.2}$^3$ & 16.51 & \phantom{0}$570^{+41}_{-37}$ & 4.54 & $11.5_{-0.4}^{+0.5}$ & M3.0 & $0.460\pm0.023$& $0.97\pm0.03$\, & $82.5\pm0.5$\, &  SEDM & 6451 ($r$), 4301 ($g$) & 2,3 \\
ZTF~J233512.60+365140.0 & 18.78 & \phantom{0}$565^{+60}_{-47}$ & 3.93 & - & M4.0 & $0.416\pm0.037$& $1.01\pm0.05$\, & $79.5\pm0.5$\,  & SEDM, DIS & 6744 ($r$), 4496 ($g$) & 2,3  \\
\hline
\\
 \multicolumn{9}{l}{Candidate polars/period-bounce polars} \\
\hline
ZTF~J011242.44+582757.1${^4}$ & 17.97 & \phantom{0}$364^{+14}_{-12}$ & 1.35 & - & late-M & $0.176 \pm 0.013$ & - & - & DBSP & - &  -  \\
\textit{ZTF~J014635.74+491443.1}${^5}$ & 16.99 & \phantom{00}$56.6^{+0.3}_{-0.3}$ & 2.06 & $8.9_{-0.5}^{+0.5}$ & BD? & - & - & 90 $\pm$ 1 & SEDM  & 5904 ($r$), 3979 &  2, 3 \\
ZTF~J034316.61-165535.8 & 19.19 & \phantom{0}$973^{+197}_{-153}$ & 1.35 & - & BD? & - & - & - & DBSP & - & - \\
ZTF~J114419.43+365747.1 & 19.91 & \phantom{0}$244^{+27}_{-24}$ & 1.82 &  $8.0_{-0.4}^{+0.4}$  & BD? & - & - & -  & SDSS & - & - \\

\hline
  \end{tabular}
  \caption{An overview of the low-accretion rate magnetic white dwarfs in PCEBs discovered or recovered by their ZTF light curves. Already known wind-accreting magnetic white dwarfs are printed in italics: $^1$: \citet{linnell2010},  $^2$: \citet{szkody2003}, $^3$: \citet{drake2014}, $^5$: \citet{guidry2021}. $^4$ has recently been mentioned by \citet{ren2023} although the identification was purely based on the ZTF light curve. 
  We include these three objects in this table because we blindly recovered them using the ZTF data and obtained a spectrum for them. The $G$-band magnitude is taken from \textit{Gaia} dr3 and the distance is taken from \citet{bailer-jones2021}. The orbital period for each object is determined from the ZTF light curve, the white dwarf temperature is determined from a fit to the SED, and the magnetic field strength is determined using equation~\ref{eq:cyclotron} based on the detected cyclotron peaks in the optical spectra. The donor mass is estimated based on the observed $K$-band magnitude and \textit{Gaia} parallax. In turn, the Roche lobe fill factor (RLFF) is based on the radius as estimated from the mass and should be interpreted with caution. }
   \label{tab:summary}
\end{table}
\end{landscape}

\begin{figure*}
    \centering
    \includegraphics{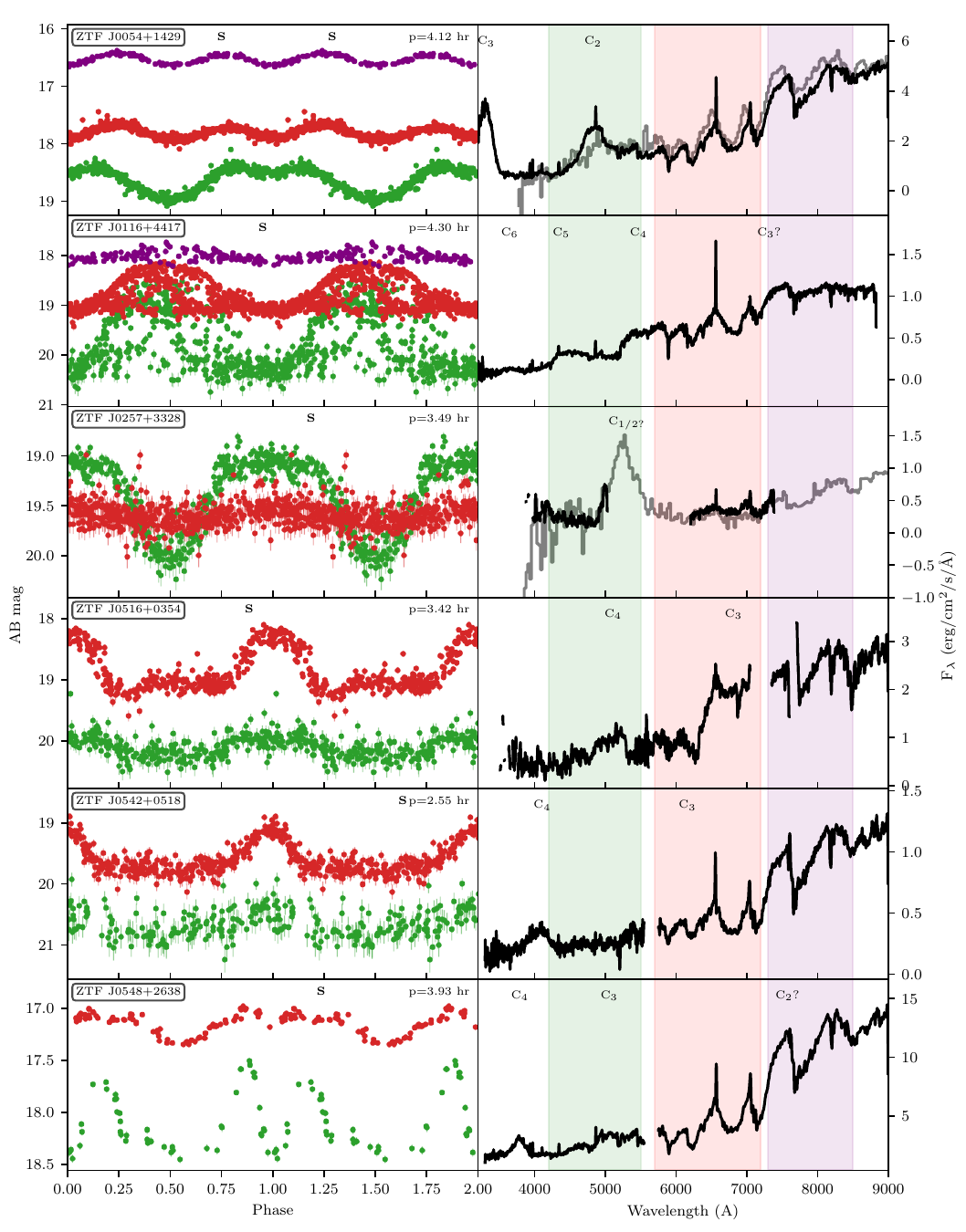}
    \caption{The left panels show the phase-folded ZTF $g$, $r$, and $i$ light curves in green, red, and purple in AB magnitudes. The right panels show one or multiple spectra (normalised). SEDM spectra are shown in grey, and other spectra are shown in black. We indicate the cyclotron harmonics with `$\mathrm{C_n}$'. The `S' in the left panel indicates at what phase the spectra were obtained. The light curve of ZTF~J0116+4417 is not strictly periodic, as shown in detail in Figure \ref{fig:lc_J0116}.}
    \label{fig:lc_spec1}
\end{figure*}
\begin{figure*}
\ContinuedFloat
    \centering
    \includegraphics{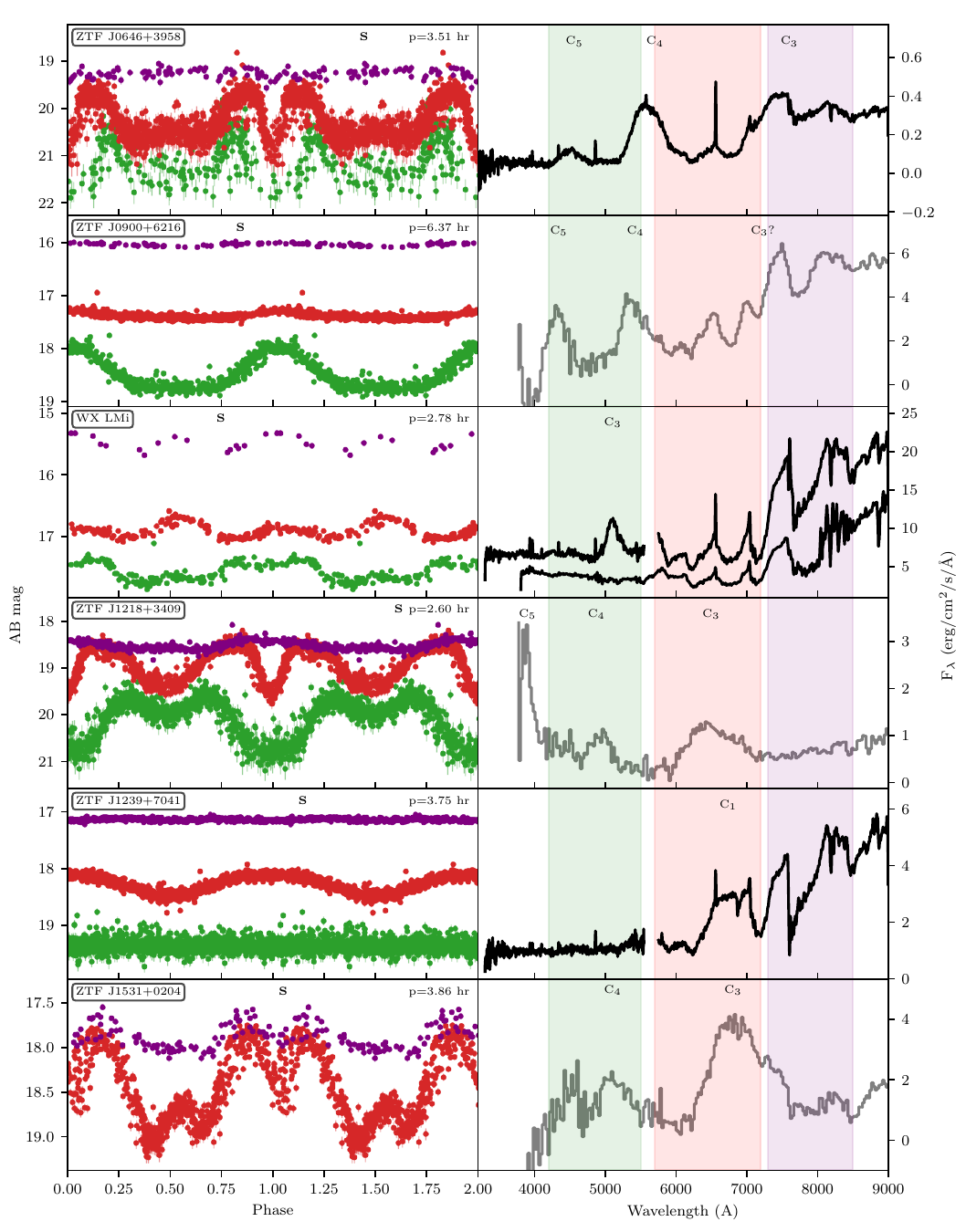}
    \caption{Continued}
    \label{fig:lc_spec2}
\end{figure*}
\begin{figure*}
\ContinuedFloat
    \centering
    \includegraphics{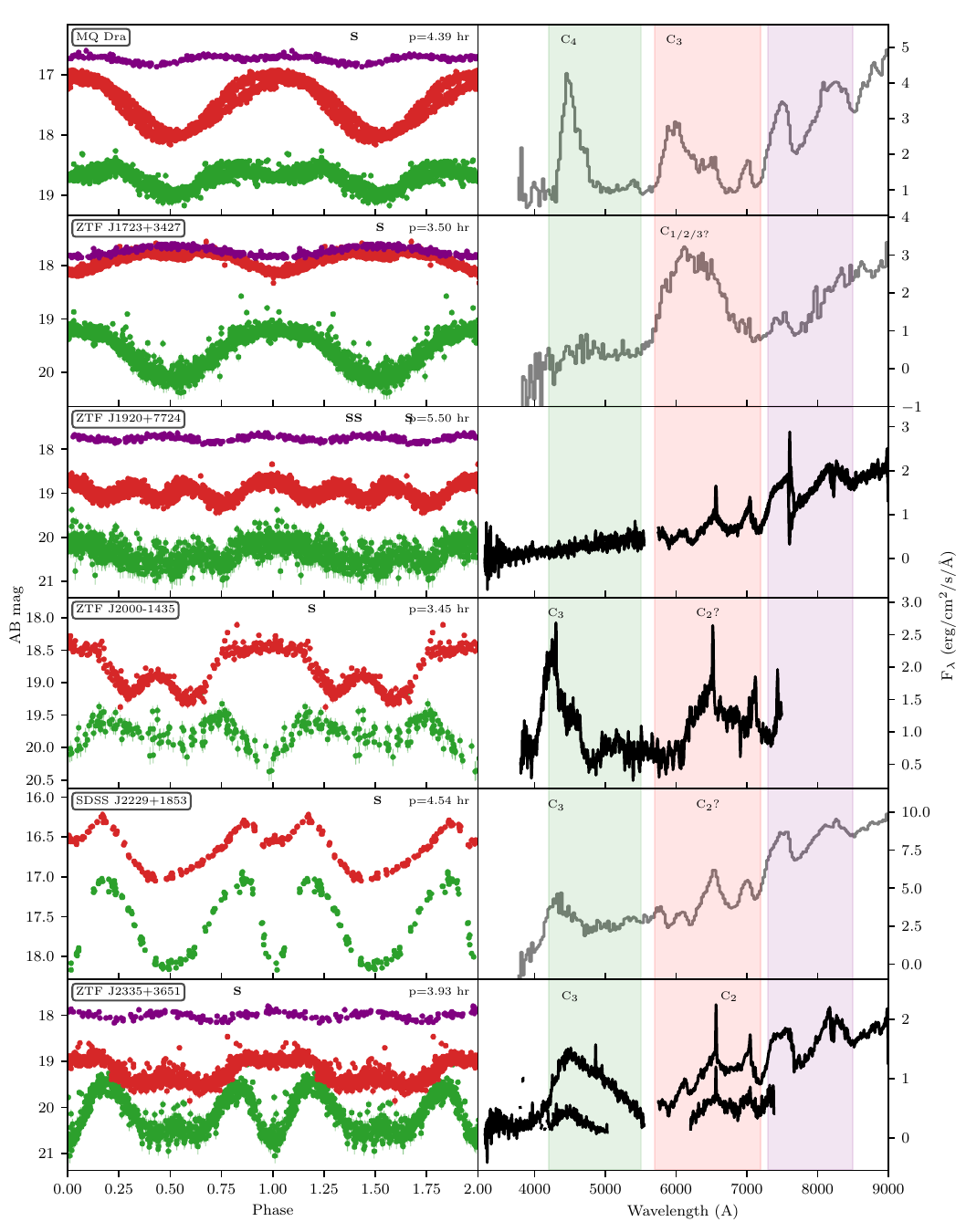}
    \caption{Continued}
    \label{fig:lc_spec3}
\end{figure*}

\begin{figure*}
    \centering
    \includegraphics{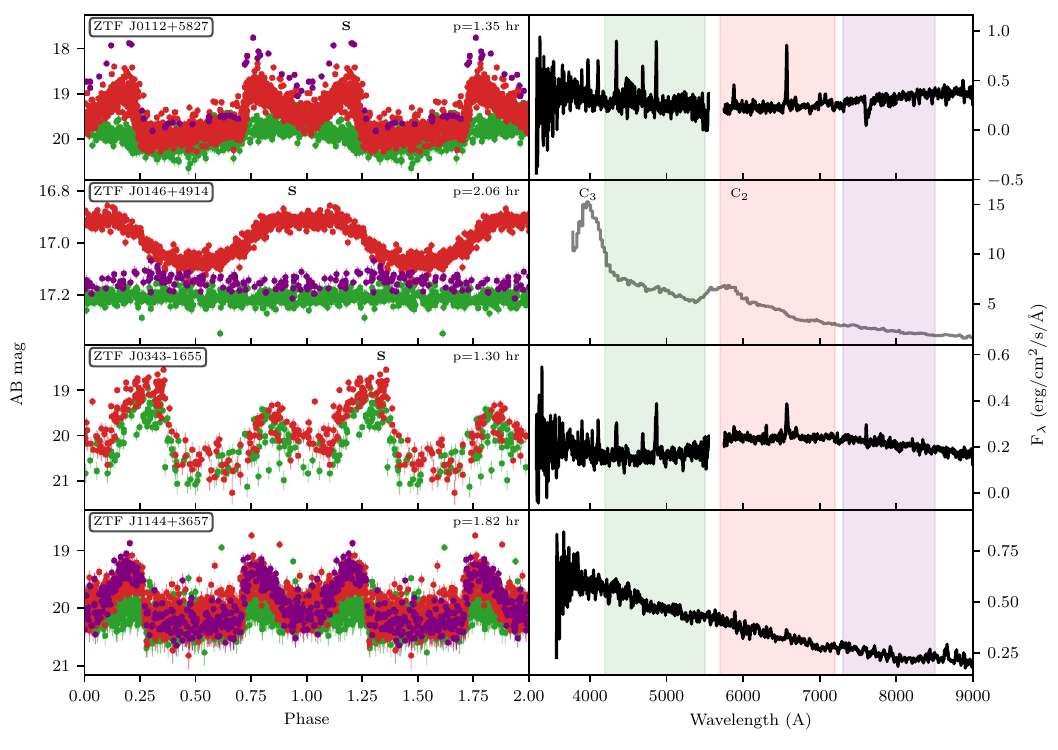}
    \caption{Similar to Fig. \ref{fig:lc_spec1}, but for objects that are likely polars. These objects passed our selection criteria (strictly periodic lightcurve, no flickering, no state-changes), but their short orbital periods and detection of a weak He-II emission line suggests that they are overfilling their Roche-lobe but have a low accretion rates and cyclotron emission is dominant.}
    \label{fig:lc_spec_polars}
\end{figure*}

\begin{figure*}
    \centering
    \includegraphics{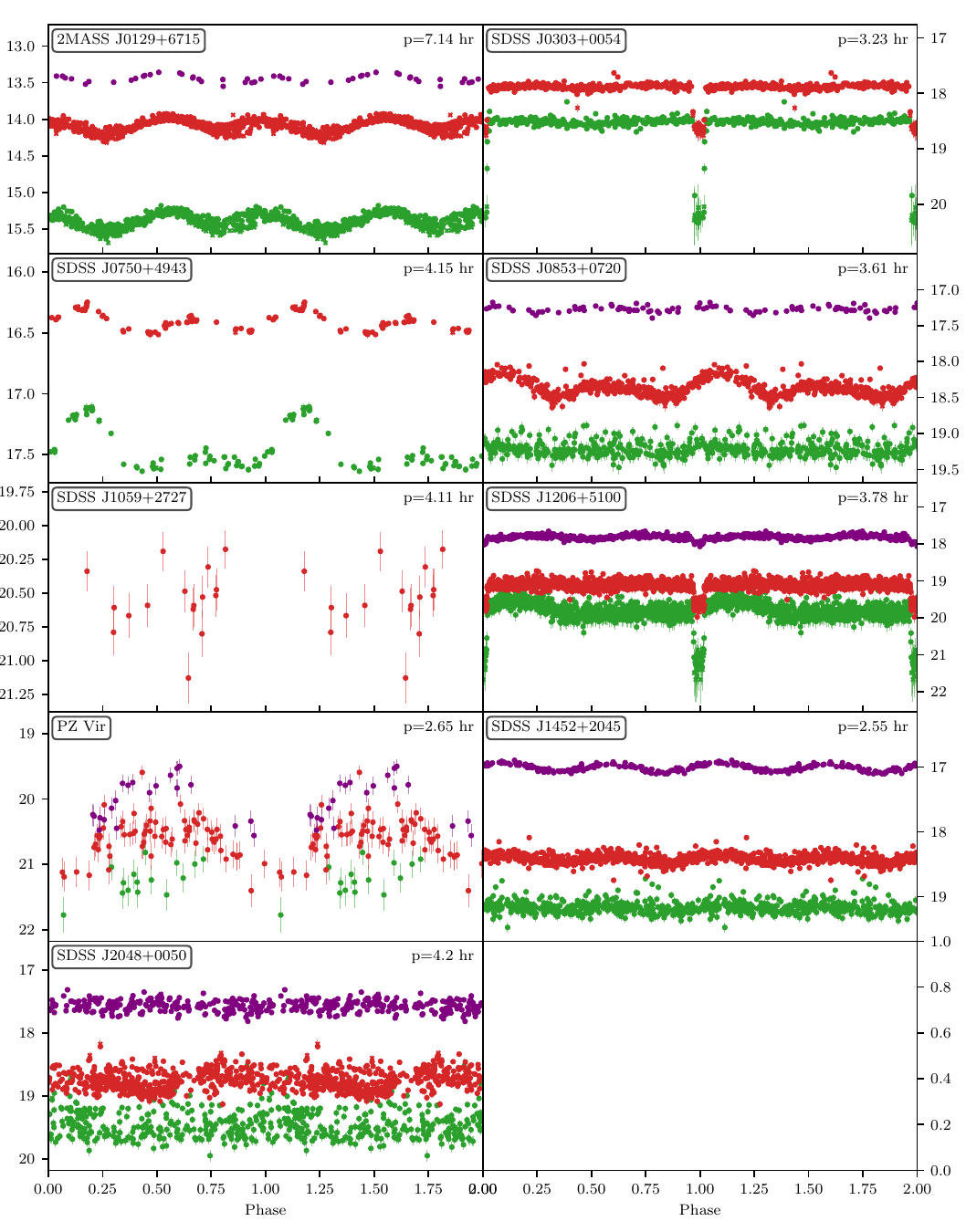}
    \caption{The ZTF light curves of known Pre-Polars as listed by \citet{parsons2021} and by \citet{hakala2022} that we did not recover in our ZTF search. }
    \label{fig:lc_known}
\end{figure*}

\begin{figure*}
    \centering
    \includegraphics{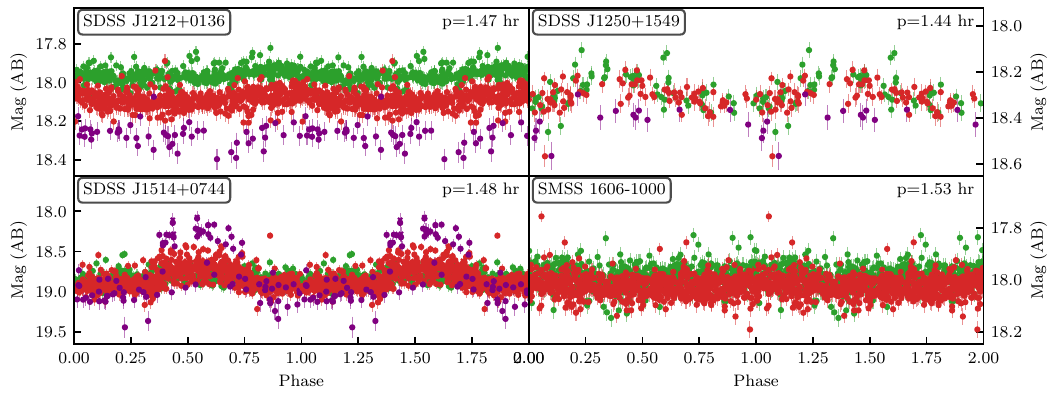}
    \caption{The ZTF light curves of four known candidate magnetic period bouncers with a low accretion rate (\citealt{breedt2012} and \citealt{kawka2021}; see \citealt{schreiber2023} for an overview). These systems only show very subtle (few percent amplitude) periodic variability that is also visible in ZTF data but not enough to identify these objects as magnetic period bouncers. }
    \label{fig:lc_known_polars}
\end{figure*}

\begin{figure*}
    \centering
    \includegraphics{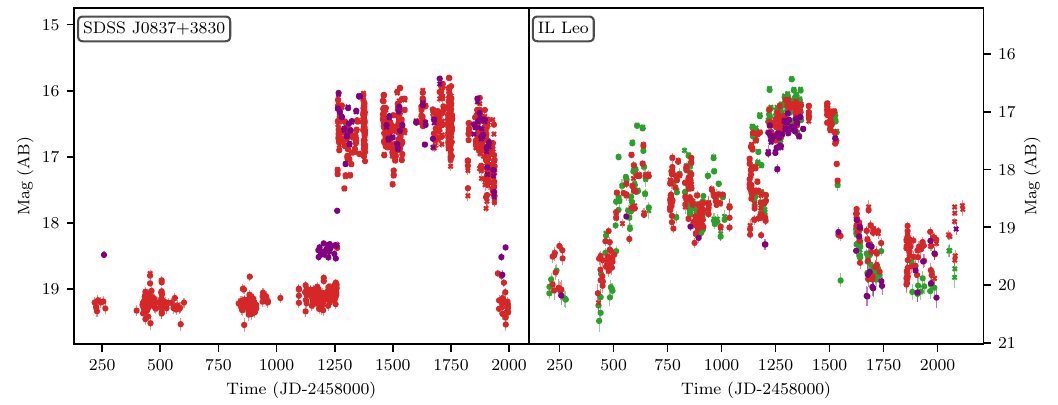}
    \caption{The ZTF light curves of SDSS J0837+3830 and IL Leo that were previously known as pre-polars. The clear change to a high state means that these objects should be reclassified as polars.}
    \label{fig:lc_known2}
\end{figure*}

\end{appendix}

\end{document}